\documentclass{article}
\usepackage{amsfonts,amssymb, amsmath,mathrsfs}

\def\on#1#2{\mathop{\vbox{\ialign{##\crcr\noalign{\kern2pt}
$\scriptstyle{#2}$\crcr\noalign{\kern2pt\nointerlineskip}
\kern-2pt$\hfil\displaystyle{#1}\hfil$\crcr}}}\limits}

\newtheorem{remark}{Remark}

\def\nn{ \nonumber }
\def\bq{ \begin{equation} }
\def\eq{ \end{equation} }
\def\ben{ \begin{eqnarray} }
\def\en{ \end{eqnarray} }
\def\frac#1#2{\genfrac{}{}{}{}{#1}{#2}}
\def\dfrac#1#2{\displaystyle{\genfrac{}{}{}{}{#1}{#2}}}

\begin{document}
%%%%%%%%%%%% TITLE %%%%%%%%%%%%%%

\title{Separation of variables for the  generalized Henon-Heiles system and system with quartic potential }

\author{Yu. A. Grigoryev and A. V. Tsiganov \\
\it\small
St.Petersburg State University, St.Petersburg, Russia\\
\it\small e--mail:
yury.grigoryev@gmail.com,  andrey.tsiganov@gmail.com}

 \date{}
\maketitle

%%%%%%%% A B S T R A C T %%%%%%%%%
\begin{abstract}
 We consider two well-known integrable systems on the plane using the concept of natural Poisson bivectors on  Riemaninan manifolds. Geometric approach to construction of  variables of separation and separated relations for the generalized Henon-Heiles system and the generalized system with quartic potential is discussed in detail.
\end{abstract}

\section{Introduction.}
\setcounter{equation}{0}

A fundamental requirement for new developments in mechanics is to unravel the geometry that
underlies different dynamical systems, especially mechanical systems. In fact,  geometric analysis of such systems reveals what they have in common and indicates the most suitable strategy to obtain and to analyze their solutions.

The Hamilton-Jacobi theory for  finite-dimensional Hamiltonian systems is well understood in both classical and geometric points of view, see foundational works of Jacobi, St\"ackel, Levi-Civita and others. Apart from  its fundamental aspects such as its relation to the action integral and generating functions of symplectic maps, the theory is known to be very useful in
 integrating the Hamilton equations using the technique of separation of
variables.

An integrable system is separable if there are variables of separation $(u,p_u)$
and $n$ separation relations
\begin{equation}
\label{seprel}
\Phi_i(u_i,p_{u_i},H_1,\dots,H_n)=0\ ,\quad i=1,\dots,n\ ,
\qquad\mbox{with }\det\left[\frac{\partial \Phi_i}{\partial H_j}\right]
\not=0\> ,
\end{equation}
connecting single pairs $u_i, p_{u_i}$ of canonical coordinates with the $n$ Hamiltonians $H_1,\ldots, H_n$.  Solving  these  relations  in terms of $p_{u_i}$  one gets the Jacobi equations and a corresponding  additively
separable complete integral of the Hamilton-Jacobi equation
\[
W=\sum_{i=1}^n \int^{u_i} p_{u_i}(u'_i, \alpha_1,\ldots,\alpha_n)\, du'_i\,,\qquad\alpha_j= H_j\,.
\]

Any  separable system is  a bi-integrable system \cite{ts07,ts10}, i.e. $n$ functionally independent integrals of motion $H_{k}$ are in  bi-involution
\bq\label{bi-inv}
\{H_i,H_k\}=\{H_i,H_k\}'=0\,,\qquad i,k=1,\ldots,n,
\eq
with respect to  compatible  Poisson brackets $\{.,.\}$ 
and   $\{.,.\}'$ associated with the Poisson bivectors $P$ and  $P'$, such that
\bq\label{comp-eq}
[P,P]=0,\qquad [P,P']=0,\qquad [P',P']=0.\qquad
\eq
Here $[.,.]$ is the  Schouten bracket.

For the given  integrable system fixed by kinematic bivector $P$ and  a tuple of integrals of motion $H_1,\ldots,H_n$    bi-Hamiltonian  construction of  variables of separation  consists in a direct solution of the equations (\ref{bi-inv}) and (\ref{comp-eq})  with respect to an unknown  bivector $P'$ \cite{ts10a,ts10k}.
The main problem is that geometrically invariant equations  (\ref{bi-inv},\ref{comp-eq}) a'priori have   infinite number  of solutions \cite{ts07,ts10}.  So, in order to get a search algorithm of  effectively computable  solutions we have to narrow the search space by using some non-invariant additional assumptions.

The aim of this note  is to prove that the concept of natural Poisson bivectors on the Riemannian manifolds allows us to properly restrict the search space and to calculate new variables of separation  for the  well-known generalized Henon-Heiles system and the system with quartic potential  \cite{gdr84,h87,rrg94}.

\section{ Settings}
\setcounter{equation}{0}
In this section we recall some  necessary facts about  natural bi-integrable  systems on Riemannian manifolds  admitting separation of variables in the Hamilton-Jacobi equation.

Let $Q$ be a $n$-dimensional  Riemannian manifold.  Its  cotangent bundle $T^*Q $ is naturally
 endowed with canonical invertible Poisson bivector $P$, which  has a standard form in fibered coordinates $z=(q_1,\ldots,q_n,p_1,\ldots,p_n)$ on  $T^*Q$
 \bq \label{p-can}
P=\left(
  \begin{array}{cc}
    0 & \mathrm{I} \\
    -\mathrm{I} & 0
      \end{array}
\right)\,,\qquad \{f,g\}=\langle P\,df,dg\rangle=\sum_{i=1}^{2n}P_{ij}\dfrac{\partial f}{\partial z_i}\dfrac{\partial g}{\partial z_j}\,.
\eq
 The Hamilton function for natural system on  $Q$
\bq \label{nat-h}
H=T+V=\sum_{i,j=1}^n \mathrm g_{ij}\,p_i\,p_j+V(q_1,\ldots, q_n)\,
\eq
is a sum of the geodesic Hamiltonian $T$ and potential energy $V$.

According to \cite{ts10,ts10s},  for the overwhelming majority of known natural Hamiltonian systems second bivector $P'$ usually  has a natural form, i.e.  $P'$ is a sum of
the geodesic Poisson bivector $P'_T$ and the  potential Poisson bivector
defined by a torsionless (1,1) tensor field $\Lambda(q_1,\ldots, q_n)$ on $Q$ associated with potential $V$
\bq\label{n-p}
P'= P'_T+\left(
      \begin{array}{cc}
        0 & \Lambda_{ij} \\
        \\
         -\Lambda_{ji}\quad &\displaystyle \sum_{k=1}^n\left(\dfrac{\partial \Lambda_{ki}}{\partial q_j}-\dfrac{\partial \Lambda_{kj}}{\partial q_i}\right)p_k
      \end{array}
    \right)\,.
\eq
The  geodesic Poisson bivector $P'_T$ is  defined by $n\times n$ matrix $\Pi$ on $T^*Q$ and functions $\mathrm{x,y}$ and $\mathrm z$:
\bq\label{p2-sph2}
 P'_T=
 \left(
      \begin{array}{cc}
        \displaystyle \sum_{k=1}^n \mathrm{x}_{jk}(q)\dfrac{\partial \Pi_{jk}}{\partial p_i}-\mathrm{y}_{ik}(q)\dfrac{\partial \Pi_{ik}}{\partial p_j} & \Pi_{ij} \\
        \\
         -\Pi_{ji}\quad&\displaystyle \sum_{k=1}^n\left(\dfrac{\partial \Pi_{ki}}{\partial q_j}-\dfrac{\partial \Pi_{kj}}{\partial q_i}\right)\,\mathrm z_{k}(p\,)\\
      \end{array}
    \right)
\,.
 \eq
In fact, functions $\mathrm{x,y}$ and $\mathrm z$ are completely determined by the matrix $\Pi$  via compatibility conditions
\bq\label{comp-pt}
[P,P'_T]=[P'_T,P'_T]=0.\eq
 We can add various integrable potentials $V$ to the given geodesic Hamiltonian $T$ in order to get integrable natural Hamiltonians (\ref{nat-h}). In  similar manner we can add different compatible potential matrices $\Lambda$ to the given geodesic matrix $\Pi$ in order to get natural Poisson bivectors $P'$ (\ref{n-p}) compatible with the canonical bivector $P$.

The definitions of natural Hamilton functions (\ref{nat-h})  and  natural Poisson bivectors (\ref{n-p}) are non-invariant because they depend on a choice  of coordinate system.  Below  we show  how  canonical transformations of variables change this definitions of natural Poisson bivectors.

\section{ Generalized Henon-Heiles system }
\setcounter{equation}{0}
Let us consider a generalized Henon-Heiles system  \cite{gdr84,h87}   defined by the following Hamilton function
 \bq\label{hh-ham}
H_1=\dfrac{p_1^2+p_2^2}{2}+\dfrac{c_1}8\,q_2(3q_1^2+16q_2^2)+c_2\left(2q_2^2
 +\dfrac{q_1^2}{8}\right)+\dfrac{c_4}{q_1^2}+\dfrac{c_5}{q_1^6}
 \eq
 and second integral of motion
 \ben
 H_2&=&p_1^4+\left(\dfrac{q_1^2(3c_1q2+c_2)}2+\dfrac{4c_4}{q_1^2}+\dfrac{4c_5}{q_1^6}\right)\,p_1^2
 -\dfrac{c_1q_1^3}{2}\,p_1p_2
 -\dfrac{c_1^2q_1^6}{32}+\dfrac{(c_2-3c_1q_2)(c_2+c_1q_2)q_1^4}{16}\nn\\
 \label{hh-sec}\\
 &+&c_1c_4q_2+\dfrac{c_2c_5+4c_4^2+3c_1c_5q_2}{q_1^4}+\dfrac{8c_4c_5}{q_1^8}+\dfrac{4c_5^2}{q_1^{12}}\,.\nn
  \en
At $c_4=c_5=0$ variables of separation  have been obtained in \cite{rrg94}, see also discussion in \cite{vz96,ts96}. In this section we recover these results in the framework of the bi-Hamiltonian geometry and obtain new variables of separation for the generic case.

 \subsection{Case $c_4=c_5=0$}
Let us suppose that the required bi-vector $P'$ has a natural form (\ref{n-p}).
Substituting   $H_{1,2}$ (\ref{hh-ham},\ref{hh-sec}) into the equation (\ref{bi-inv}) and solving resulting  equation together with (\ref{comp-eq})  one gets two distinct solutions $P'_1$ and $P'_2$.

First solution  $P'_1$ is defined by the following geodesic matrix
\bq\label{pi-1}
 \Pi^{(1)}=\dfrac{1}{2}\left(
      \begin{array}{cc}
        p_1^2+\frac{1}{2}p_2^2 & 0 \\ \\
        \frac{1}{2}p_1p_2 & \frac{1}{2}p_2^2
      \end{array}
    \right)\eq
and potential matrices
 \bq
\Lambda^{(1)}=\left(
          \begin{array}{cc}
            \frac{q_1^2(3c_1q_2+c_2)}{8}+c_1q_2^3+c_2q_2^2\qquad &
             \frac{c_1q_1^3}{16}+\left(\frac{3c_1q_2}{2}+c_2\right)q_1q_2
             \\ \\
            -\frac{c_1q_1^3 }{32}& c_1q_2^3+c_2q_2^2
          \end{array}
        \right)\,,
     \eq
for which
\[\mathrm x_{jk}=\mathrm y_{jk}=\delta_{jk}q_k\,,\qquad z_k(p)=0.\]
 The corresponing  recursion operator $N_1=P'_1P^{-1}$ yields the integrals of motion
\begin{equation}\label{aux-int}
\mathcal{H}_k=\frac1{2k}\,\mathrm{tr}\,N_1^k\,,\qquad k=1,2,
\end{equation}
which form a bi-Hamiltonian hierarchy, i.e. the Lenard relations hold
\bq\label{len-rel}
P' d\mathcal{H}_1=P d{\mathcal H}_{2}. \eq
These integrals are the following functions of  initial integrals of motion (\ref{hh-ham},\ref{hh-sec})
\[\label{hh-aux}
 \mathcal{H}_1=2H_1\qquad\qquad \mathcal{H}_2=\dfrac{H_2}{8}+\dfrac{H_1^2}{2}\,.
\]
 So,  the  first recursion operator $N_1=P'_2P^{-1}$ gives rise to the  action variables $\mathcal H_{1,2}$ (\ref{aux-int}) with trivial dynamics.

Second solution $P'_2$ is defined by the matrices
\bq\label{pi-2}
 \Pi^{(2)}=\dfrac{1}{2q_1^2}\left(
      \begin{array}{cc}
       2p_1^2 & 0 \\ \\
              p_1p_2&  0
      \end{array}
    \right)\,,\qquad
\Lambda^{(2)}=\left(
                \begin{array}{cc}
                \dfrac{c_1q_2}{2}+\dfrac{c_2}{4}\quad& \dfrac{c_1q_1}{8}+\dfrac{q_2(3c_1q_2+2c_2)}{q_1}\\ \\
                -\dfrac{c_1q_1}{16}&-\dfrac{c_1q_2}{4}\end{array}
              \right)\,
\eq
and functions
\[
\mathrm x_{j1}=\mathrm y_{i1}=-q_1\,,\qquad \mathrm  z_k(p)=0\,.
\]
Instead of the Lenard relations (\ref{len-rel}) here we have the following relations
\bq\label{f-mat}
P'dH_i=P\sum_{j=1}^2 F_{ij}\,dH_j,\qquad i=1,\ldots,2,
 \eq
where $F$ is a so-called control matrix
\[
F=\dfrac{1}{2}\left(
    \begin{array}{cc}
     4p_1^2q_1^{-2}+c_1q_2+c_2\quad & q_1^2 \\ \\
     16H_2\,q_1^{-2}  & 4p_1^2q_1^{-2}+c_1q_2+c_2
    \end{array}
  \right)\,.
\]
Eigenvalues of $F$ coincide  with the eigenvalues of  recursion operator, which are the desired  variables of separation.

So, the second recursion operator $N_2=P'_2P^{-1}$ yields the variables of separation  with non-trivial se\-pa\-ra\-ted relations. These variables of separation and the corresponding  separated relations are considered in the next section.

As usual, the  recursion operators $N_{1,2}$ generate two infinite  families of solutions of the equations (\ref{bi-inv},\ref{comp-eq})
\bq\label{h-poi}
P_{1,2}^{(m)}=N_{1,2}^m \,P\,,\qquad m=\ldots,-1,0,1,\ldots,
\eq
associated with the  Hamiltonians $H_{1,2}$ (\ref{hh-ham},\ref{hh-sec}).

 \subsection{Case $c_{4,5}\neq0$}
 According to \cite{ts10} at $c_{5}\neq0$  we have to apply the following canonical transformation
\bq\label{ctr-hh}
 p_1\to p_1+\sqrt{\dfrac{-2c_5}{q_1^6}}\,,
 \eq
 to the natural Poisson bivectors $P'_{1,2}$.   Namely, in generic case the  integrals of motion $H_{1,2}$ (\ref{hh-ham},\ref{hh-sec}) are in involution with respect to the Poisson brackets associated with the shifted bivectors
\bq\label{hhg-p1}
 \hat{P}_1=P'_1+\dfrac{\sqrt{-2c_5}}{q_1^3}\left(
                  \begin{array}{cccc}
                    0 &  0& p_1+\dfrac{1}{2}\sqrt{\dfrac{-2c_5}{q_1^6}} &0  \\
                    * & 0 & p_2 & 0 \\
                    *& * & 0 & \dfrac{3c_1}{8}q_1^2+6c_1q_2^2+4c_2q_2 \\
                    * & * & * & 0 \\
                  \end{array}
                \right)
 \eq
 and
\bq\label{hhg-p2}
\hat{P}_2=P'_2+\dfrac{2 \sqrt{-2c_5}}{q_1^5}\left(
                  \begin{array}{cccc}
                    0 & 0 &p_1+\dfrac{1}{2}\sqrt{\dfrac{-2c_5}{q_1^6}}  & 0 \\
                    * & 0 & {p_2}& 0 \\
                    * &* & 0 & \dfrac{3c_1q_1^2}{8}+6c_1q_2^2+4c_2q_2 \\
                    * & * & * & 0 \\
                  \end{array}
                \right)\,.
\eq
These bivectors were obtained by substituting non-homogeneous polynomial ansatz for the geo\-de\-sic bivector $P'_T$ into the equation  (\ref{comp-pt}). In contrast with the pair of  bivectors $P'_{1,2}$ (\ref{pi-1},\ref{pi-2})  these  bivectors $\hat{P}_{1,2}$ generate variables of separation with nontrivial separated relations in  both cases.

 Firstly, let us consider the recursion operator $\hat{N}_2=\hat{P}_2P^{-1}$ and its eigenvalues $u_{1,2}$, which are the roots of the following  polynomial
\ben
{B}(\lambda)&=&(\lambda-u_1)(\lambda-u_2)=\lambda^2-\left(\dfrac{p_1^2}{q_1^2}+\dfrac{c_1q_2+c_2}{4}+\dfrac{2\sqrt{-2c_5}\,p_1}{q_1^5}-\dfrac{2c_5}{q_1^8}\right)\lambda
\nn\\
\label{hh-u}\\
&-&\dfrac{c_1(4p_1^2q_2-2q_1p_1p_2-c_2q_1^2q_2)}{16q_1^2}+\dfrac{c_1^2(8q_1^2+q_2)}{16}
-\dfrac{c_1\sqrt{-2c_5}(4p_1q_2-q_1p_2)}{8q_1^5}
+\dfrac{c_1c_5q_2}{2q_1^8}\,.\nn
\en
According to \cite{ts10a,ts10k,ts10} now we are looking for the auxiliary polynomial $A(\lambda)=a_1\lambda+a_0$, which  is  solution of the equation
\bq\label{ab-eq}
\{B(\lambda),A(\mu)\}=-\dfrac{\left(d_2\mu^2+d_1\mu+d_0\right)B(\lambda)-\left(d_2\lambda^2+d_1\mu+d_0\right)B(\mu)}{\lambda-\mu}\,,\qquad
\{A(\lambda),A(\mu)\}=0
\eq
with respect to unknown  functions $a_{1,0}$, $d_{1,2}$ and $d_0$. In our  case $d_2=d_0=0$,  $d_0=1$, and desired polynomial
\[
 A(\lambda)=-\dfrac{64(q_1^3p_1+\sqrt{-2c_5})}{c_1^2q_1^4}\,\lambda-\dfrac{4(4p_1q_2-q_1p_2)}{c_1q_1}
-\dfrac{16\sqrt{-2c_5}q_2}{c_1q_1^4}
\]
satisfies the following equations
\[
\{{B}(\lambda), A(\mu)\}=-\dfrac{1}{\lambda-\mu}\,\Bigl({B}(\lambda)-{B}(\mu)\Bigr)\,,\qquad \{A(\lambda),A(\mu)\}=0\,.
\]
It entails that
\[
p_{u_j}=A(\lambda=u_j)\,,\qquad \{u_i,p_{u_j}\}=\delta_{ij}\,,\qquad i,j=1,2,
\]
are  canonically conjugated momenta.

An inverse canonical transformation from variables of separation to  the initial variables looks like
\ben
q_1&=&\sqrt{\dfrac{c_1^2(p_{u_1}^2-p_{u_2}^2)}{32\,(u_1-u_2)}
+32\,\dfrac{c_2(u_1+u_2)-4(u_1^2-u_1u_2-u_2^2)}{c_1^2}}\,,\nn\\
\nn\\
p_1&=&-\dfrac{c_1^2(p_{u_1}-p_{u_2})}{64(u_1-u_2)}\,q_1-\dfrac{\sqrt{-2c_5}}{q_1^3}\,,\qquad \qquad
q_2=-c_1^3\left(\dfrac{p_{u_1}-p_{u_2}}{32(u_1-u_2)}\right)^2+\dfrac{4(u_1+u_2)-c_2}{c_1}
\,,\nn\\
\nn\\
p_2&=&2c_1^5\left(\dfrac{p_{u_1}-p_{u_2}}{32(u_1-u_2)}\right)^3
+\dfrac{c_1}{4(u_1-u_2)}\left(
\dfrac{p_{u_1}-p_{u_2}}{4}\,c_2
-(u_1+2u_2)p_{u_1}+(2u_1+u_2)p_{u_2}\right)
\,.\nn
\en
Now it is easy to calculate  the  separated relations
\bq\label{hh-seprel}
\Phi(u_k,p_{u_k})=\Phi_+(u_k,p_{u_k})\Phi_-(u_k,p_{u_k}) -\frac{c_4(c_2-8u_k)}{4}
+\frac{c_1^2\sqrt{-2c_5}\,p_{u_k}}{32}=0\,,\qquad k=1,2,
\eq
where
\[
\Phi_\pm(u_k,p_{u_k})=
\left(\frac{c_1^2p_{u_k}^2}{32}-H_1\pm\frac{\sqrt{H_2}}{2}-\frac{128u_k^3}{c_1^2}+\frac{32c_2u_k^2}{c_1^2}\right)\,.
\]
These separated relations are given by the affine equations in Hamiltonians  $H_1$ and $H_2-4H_1^2$.
It means that the generalized Henon-Heiles system  belongs to the St\"ackel  family of separable systems.

At $c_4=c_5=0$ these separated relations (\ref{hh-seprel}) may be  reduced to a pair of distinct separated  relations
\[\Phi_+(u_1,p_{u_1})=0\,,\qquad\mbox{and}\qquad \Phi_-(u_2,p_{u_2})=0\,.\]
and the  equations of motion are linearized on  two  different elliptic curves, see \cite{vz96,rrg94,ts96}.

 In generic case $c_4\neq0$ and $c_5\neq 0$ equations of motion are linearized on the two copies of non-hyperelliptic curve of genus three defined by  (\ref{hh-seprel}).  An explicit description of the   linerization procedure is an open problem similar to  the generalized  Kowalevski and Chaplygin systems \cite{ts10a,ts10k}.

Using the first bivector $\hat{P}_1$ (\ref{hhg-p1}) we can get other variables of separation
\[
\widehat{B}(\lambda)=\Bigl(\,\det(\hat{P}_1P^{-1}-\lambda\mathrm I)\,\Bigr)^{1/2}=(\lambda-v_1)(\lambda-v_2)\,,
\]
which are related with the previous separated variables by canonical transformation
\[
v_k=\dfrac{c_1^2p_{u_k}^2}{64}-\dfrac{64 u_k^3}{c_1^2}+\dfrac{16c_2u_k^2}{c_1^2}\,,\qquad k=1,2.
\]

\section{ Generalized system with quartic potential}

\setcounter{equation}{0}

Let us consider a generalization of the system with quartic potential \cite{h87}  with  the Hamilton function
\begin{equation}
\label{qp-h1}
H_1=\frac{p_1^2+p_2^2}{2} + \frac{c_1}{4}\left(q_1^4+6q_1^2q_2^2+8q_2^4\right)+\frac{c_2}{2}\left(q_1^2+4q_2^2\right)+\frac{2c_3}{q_2^2}+\frac{c_4}{q_1^2}+\frac{c_5}{q_1^6}
\end{equation}
and second integral of motion
\begin{eqnarray}
\label{qp-h2}
H_2& = p_1^4&+p_1^2\left(c_1q_1^4+6c_1q_1^2q_2^2+2c_2q_1^2+\frac{4c_4}{q_1^2}+\frac{4c_5}{q_1^6}\right)-4c_1q_1^3q_2p_1p_2\nonumber\\
&&+c_1q_1^4p_2^2+\frac{4c_4^2}{q_1^4}+2c_1c_4q_1^2+4c_1c_4q_2^2+c_2^2q_1^4+c_1c_2q_1^6+2c_1c_2q_1^4q_2^2+\frac{c_1^2q_1^8}{4}\nonumber\\
&&+c_1^2q_1^6q_2^2+c_1^2q_1^4q_2^4+\frac{4c_1c_3q_1^4}{q_2^2}+c_5\left(\frac{8c_4}{q_1^8}+\frac{4c_5}{q_1^{12}}+\frac{4c_2}{q_1^4}+\frac{2c_1}{q_1^2}+\frac{12c_1}{q_1^4q_2^2}\right)\,.
\end{eqnarray}
At  $c_4=c_5=0$ variables of separation were obtained in \cite{rrg94}, see also \cite{ts96}.
 In the following section we  reproduce this result in the framework of bi-Hamiltonian geometry and obtain variables of separation in the generic case at $c_{4,5}\ne0$.

\subsection{Case  $c_4=c_5=0$}

According to  \cite{ts06,ts10}, the geodesic matrices $ \Pi^{(1,2)}$ (\ref{pi-1},\ref{pi-2})
are compatible  with another pair of   potential matrices
\bq\label{qq-lambda1}
\Lambda^{(1')}=\left(
          \begin{array}{cc}
           \dfrac{c_1q_1^4}{4}+\left(3c_1q_2^2+c_2\right)\dfrac{q_1^2}{2}
           +c_1q_2^4+c_2q_2^2+\dfrac{c_3}{q_2^2}  \quad &\dfrac{c_1q_1^3q_2}{2}+\left(2c_1q_2^3+c_2q_2-\dfrac{c_3}{q_2^3}\right)q_1
             \\ \\
             -\dfrac{c_1q_1^3q_2}{4}& c_1q_2^4+c_2q_2^2+\dfrac{c_3}{q_2^2}
          \end{array}
        \right)
\eq
and
\begin{equation}
\label{qq-lambda2}
\Lambda^{(2')}=\left( \begin{array}{cc}
c_2+c_1\left(\dfrac{q_1^2}{2}+2q_2^2\right) \quad& c_1q_1q_2+2\dfrac{2c_1q_2^6+c_2q_2^4-c_3}{q_1q_2^3} \\
-\dfrac{c_1q_1q_2}{2} & -c_1q_2^2 \end{array} \right)\,.
\end{equation}
As above,  a natural  Poisson bivector $P'_1$ defined by the matrices $ \Pi^{(1)}$ and $ \Lambda^{(1')}$ yields recursion operator $N_1$ and integrals of motion  $\mathcal H_{1,2}$ (\ref{aux-int}), which form a bi-Hamiltonian hierarchy (\ref{len-rel}).

Matrices $ \Pi^{(2)}$ and $ \Lambda^{(2')}$ define the second natural bivector $P'_2$ and
the  recursion operator $N_2=P'_2P^{-1}$, whose eigenvalues are the variables of separation  $u_{1,2}$
\begin{eqnarray}
\label{Bpoly}
B(\lambda)&=&\Bigl(\,\det(N_2-\lambda\mathrm I)\,\Bigr)^{1/2}=(\lambda-u_1)(\lambda-u_2)\\
&=&\lambda^2+\dfrac{q_1^4c_1+2c_2q_1^2+2q_1^2c_1q_2^2+2p_1^2}{2q_1^2}\,\lambda
+\dfrac{c_1(q_2^2q_1^2p_2^2+4c_3q_1^2+4p_1^2q_2^4-4p_1q_2^3q_1p_2)}{4q_1^2q_2^2}\,.\nn
\end{eqnarray}
In this case equation (\ref{ab-eq}) has another  solution  $d_2=d_0=0$, $d_1=1$, and
\begin{equation}
\label{Apoly}
A(\lambda)=-\frac{p_1}{c_1q_1}\,\lambda-\frac{q_2^2p_1}{q_1}+\frac{q_2p_2}{2}\,.
\end{equation}
It  allows us to  calculate  the desired momenta
\begin{equation}
\label{pfromA}
p_{u_j}=\frac{A(u_j)}{u_j}\,,\qquad \{u_i,p_{u_j}\}=\delta_{ij}\,,\qquad i,j=1,2.
\end{equation}
In the variables of separation $(u,p_u)$ the second integral of motion  $H_2$ (\ref{qp-h2}) is a complete square and it is easy to prove that $H_1$ and $\sqrt{H_2}$  satisfy the following separated relations
\begin{eqnarray}
\Phi_-\left(u,p_{u}\right)&=H_1-\dfrac{1}{2}\sqrt{H_2}+2c_1up_{u}^2-\dfrac{2u^2}{c_1}
+\dfrac{2c_2u}{c_1}+\dfrac{2c_1c_3}{u}=0\,,\qquad &u=u_1,~p_u=p_{u_1}\,,\nonumber\\
\label{Phis}\\
\Phi_+\left(u,p_{u}\right)&=H_1+\dfrac{1}{2}\sqrt{H_2}+2c_1up_{u}^2-\dfrac{2u^2}{c_1}
+\dfrac{2c_2u}{c_1}+\dfrac{2c_1c_3}{u}=0\,, \qquad &u=u_2,~p_u=p_{u_2}\,,\nonumber
\end{eqnarray}
and  equations of motion are linearized on a pair of elliptic curves (see \cite{vz96,rrg94,ts96}).

\subsection{Case $c_{4,5}\neq0$}
 At $c_{5}\neq0$ in order to get bi-Hamiltonian structures we apply the same  canonical transformation
as for the Henon-Heiles system (\ref{ctr-hh})
\bq\label{ctr-qq}
 p_1\to p_1+\sqrt{\dfrac{-2c_5}{q_1^6}}\,,
 \eq
 which shifts  the  Poisson bivectors defined by potential matrices (\ref{qq-lambda1},\ref{qq-lambda2})  by the rule
\bq\label{qqg-p1}
\widetilde{P}_1=P'_1+\dfrac{ \sqrt{-2c_5}}{q_1^3}\left(
                  \begin{array}{cccc}
                    0 &  0& p_1+\dfrac{1}{2}\sqrt{\dfrac{-2c_5}{2q_1^6}} & 0 \\
                    * & 0 & p_2 & 0 \\
                    *& * & 0 &8c_1q_2^3+(3c_1q_1^2+4c_2)q_2-\dfrac{4c_3}{q_2^3} \\
                    * & * & * & 0 \\
                  \end{array}
                \right)
\eq
 and
\bq\label{qqg-p2}
\widetilde{P}_2=P'_2+\dfrac{ 2\sqrt{-2c_5}}{q_1^5}\left(
                  \begin{array}{cccc}
                    0 & 0 &p_1+\dfrac{1}{2}\sqrt{\dfrac{-2c_5}{q_1^6}}  & 0 \\
                    * & 0 & {p_2}& 0 \\
                    * &* & 0 &8c_1q_2^3+(3c_1q_1^2+4c_2)q_2 -\dfrac{4c_3}{q_2^3}\\
                    * & * & * & 0 \\
                  \end{array}
                \right)\,.
\eq
 At $c_5\neq0$ polynomials $B(\lambda)$ and $A(\lambda)$ are obtained from  (\ref{Bpoly})  and (\ref{Apoly}) using the canonical transformation (\ref{ctr-qq}). For instance, a  "shifted'' polynomial $A(\lambda)$ reads as
\[
A(\lambda)=-\dfrac{\left(p_1+\sqrt{\dfrac{-2c_5}{q_1^6}}\right)\lambda}{c_1q_1}-\dfrac{q_2^2\left(p_1+\sqrt{\dfrac{-2c_5}{q_1^6}}\right)}{q_1^2}+\dfrac{p_2q_2}{2}\,.
\]
An inverse canonical transformation from variables of separation to the original variables looks like
\ben
q_1&=&\sqrt{
\left(-\dfrac{2(u_1p_{u_1}^2-u_2p_{u_2}^2)}{u_1-u_2}+\dfrac{2c_3}{u_1u_2}\right)c_1+\dfrac{2(u_1+u_2-c_2)}{c_1}
}\,,
\nn\\
p_1&=&\dfrac{c_1(u_1p_{u_1}-u_2p_{u_2})}{u_1-u_2}\,q_1-\dfrac{\sqrt{-2c_5}}{q_1^3}\,,\qquad\qquad
q_2=\sqrt{
-\dfrac{c_1u_1u_2(p_{u_1}-p_{u_2})^2}{(u_1-u_2)^2}-\dfrac{c_3}{u_1u_2}
}\,,
\nn\\
p_2&=&-\dfrac{2}{q_2}\left(\dfrac{c_1^2c_3(u_1p_{u_1}-u_2p_{u_2})}{u_1u_2(u_1-u_2)}
+\dfrac{u_1u_2(p_{u_1}-p_{u_2})}{(u_1-u_2)^3}\Bigl(c_1^2(p_{u_1}-p_{u_2})(u_1p_{u_1}-u_2p_{u_2})-(u_1-u_2)^2\Bigr)
\right)\,.\nn
\en
Now it is easy to prove that the corresponding separated relations are equal to
\bq\label{qq-seprel}
\Phi(u_k,p_{u_k})=\Phi_+(u_k,p_{u_k})\Phi_-(u_k,p_{u_k}) +(2u_k-c_2)c_4-2c_1\sqrt{-2c_5}\,u_k\,p_{u_k}=0\,,\qquad k=1,2,
\eq
where $\Phi_{\pm}(u,p_u)$ are given by (\ref{Phis}).  The separated relations are  affine equations with respect to  the Hamilton functions  $H_1$ and integral of motion  $H_2-4H_1^2$. It means that the generalized system with quartic potential  belongs to the St\"ackel  family of separable systems.

As above, in the generic case $c_4\neq0$ and $c_5\neq 0$ equations of motion are linearized on  the two copies of the non-hyperelliptic curve of genus three defined by  (\ref{qq-seprel}) and we do not know how to solve the corresponding Abel-Jacobi equations as yet.

Using the first bivector $\tilde{P}_1$ (\ref{qqg-p1}) we can get other variables of separation
\[
\widetilde{B}(\lambda)=\Bigl(\,\det(\widetilde{P}_1P^{-1}-\lambda\mathrm I)\,\Bigr)^{1/2}=(\lambda-v_1)(\lambda-v_2)\,,
\]
which are related with  previous separated variables by the following  canonical transformation
\[
v_k=c_1u_kp_{u_k}^2-\dfrac{u_k^2}{c_1}+\dfrac{c_2u_k}{c_1}+\dfrac{c_1c_3}{u_k}\,,\qquad k=1,2.
\]

\end{document}